\documentclass[apj]{emulateapj}
\usepackage{amsmath}

\usepackage{epsf}

\shortauthors{Zucker et al.}
\shorttitle{Andromeda X}

\newcommand{\aix}{And~IX}
\newcommand{\ax}{And~X}

\begin{document}

\title{Andromeda X, A New Dwarf Spheroidal Satellite of M31: Photometry}

\author{Daniel B.\ Zucker\altaffilmark{1,2}, 
Alexei Y.\ Kniazev\altaffilmark{1,3}, 
David Mart\'{i}nez-Delgado\altaffilmark{1,4},
Eric F.\ Bell\altaffilmark{1},   
Hans-Walter Rix\altaffilmark{1}, 
Eva K.\ Grebel\altaffilmark{5},  
Jon A.\ Holtzman\altaffilmark{6}, 
Rene A.\ M.\ Walterbos\altaffilmark{6},
Constance M.\ Rockosi\altaffilmark{7},
Donald G. York\altaffilmark{8},
J. C. Barentine\altaffilmark{9},
Howard Brewington\altaffilmark{9},
J.~Brinkmann\altaffilmark{8},
Michael Harvanek\altaffilmark{9},
S.\ J.\ Kleinman\altaffilmark{9},
Jurek Krzesinski\altaffilmark{9,10},
Dan Long\altaffilmark{9},
Eric H. Neilsen, Jr.\altaffilmark{11},
Atsuko Nitta\altaffilmark{9},
Stephanie A.\ Snedden\altaffilmark{9}
}

\altaffiltext{1}{Max-Planck-Institut f\"ur Astronomie,
K\"onigstuhl 17, D-69117 Heidelberg, Germany}
\altaffiltext{2}{Current address: Institute of Astronomy, Madingley Road, Cambridge, CB3 0HA, United Kingdom; \texttt{zucker@ast.cam.ac.uk}}
\altaffiltext{3}{South African Astronomical Observatory, PO Box 9, Observatory 7935, Cape Town, South Africa}
\altaffiltext{4}{Instituto de Astrof\'{\i}sica de Andaluc\'{\i}a (CSIC), C/Camino Bajo de Hu\'{e}tor 50, 18008 Granada, Spain}
\altaffiltext{5}{Astronomisches Institut, Universit\"{a}t Basel, Venusstrasse 7,CH-4102 Binningen, Switzerland}
\altaffiltext{6}{Department of Astronomy, New Mexico State University,
1320 Frenger Mall, Las Cruces NM 88003-8001}
\altaffiltext{7}{UCO/Lick Observatory, 1156 High St., Santa Cruz, CA 95064}
\altaffiltext{8}{Department of Astronomy and Astrophysics, University of Chicago, 5640 South Ellis Avenue, Chicago, IL 60637}
\altaffiltext{9}{Apache Point Observatory, P.O. Box 59, Sunspot, NM 88349}
\altaffiltext{10}{Mt. Suhora Observatory, Cracow Pedagogical University, ul. Podchorazych 2, 30-084 Cracow, Poland}
\altaffiltext{11}{Fermi National Accelerator Laboratory, P.O. Box 500, Batavia, IL 60510}

\begin{abstract}

We report the discovery of Andromeda~X, a new dwarf spheroidal satellite of M31, based on stellar photometry from the Sloan Digital Sky Survey (SDSS).  Using follow-up imaging data we have estimated its distance and other physical properties. We find that Andromeda~X has a dereddened central surface brightness of $\mu_{V,0} \sim 26.7$ mag arcsec$^{-2}$ and a total apparent magnitude of $V_{\rm tot} \sim 16.1$, which at the derived distance modulus, $(m - M)_0 \sim 24.12 - 24.34$, yields an absolute magnitude of $M_V \sim -8.1 \pm 0.5$; these values are quite comparable to those of Andromeda~IX, a previously-discovered low luminosity M31 satellite. The discoveries of Andromeda~IX and Andromeda~X suggest that such extremely faint satellites may be plentiful in the Local Group. 
\end{abstract}

\keywords{galaxies: dwarf  --- galaxies: individual (Andromeda IX, Andromeda X) --- galaxies: evolution --- Local Group}

\section{Introduction}
\label{txt:intro}

In hierarchical cold dark matter (CDM) models, large galaxies like the Milky Way and
M31 form from the merger and accretion of smaller systems. Such models, while successful at large scales, predict at least 1 -- 2 orders of magnitude more low-mass dark subhalos at the present epoch than the observed abundance of dwarf galaxies \citep[e.g.,][]{klyp99,moor99,bens02b}. This discrepancy, the ``missing satellite'' problem, is one of the most serious obstacles for matching CDM theory to observations.

A number of solutions have been proposed to address the problem, at least qualitatively.
Star formation in low mass subsystems could be inhibited
\citep[e.g.,][]{some02,bens02a}. It is possible that the observed satellites are in fact embedded in much larger, more massive dark subhalos \citep{stoe02}, or originally formed in more massive subhalos that were later tidally stripped \citep{krav04}, and that we are therefore observationally sampling a less populous region of the initial satellite mass function.

All of these solutions aim to resolve the discrepancy between theory and observation by creating models which predict fewer directly observable satellites. A complementary observational approach would be to place more stringent constraints on the faint end of the galaxy luminosity function, but attempts to do so are hampered by the low surface brightnesses expected of galaxies in that regime.
The advent of wide-field CCD surveys such as the Sloan Digital Sky Survey (SDSS) has made it possible to detect stellar structures with extremely low surface brightnesses \citep[see, e.g.,][]{will02},
and in the past two years data from SDSS have been used to find two new 
Local Group members, Andromeda IX \citep[\aix;][]{zuck04b} and Ursa Major \citep{will05}, satellites of M31 and the Milky Way, respectively; at the time of its discovery, each of these two objects was determined to be the least luminous, lowest surface brightness galaxy found up to that point.

Since the discovery of \aix, we have been using SDSS photometry of M31 and its surroundings to select fields for deeper observations on other telescopes, in order to search for additional M31 companions. In this letter, we report the discovery, using SDSS and followup photometric data, of a new dwarf spheroidal companion to M31, one which appears comparable in luminosity to \aix. Keck spectroscopy of stars in the field of this new satellite, and the kinematic and abundance information derived therefrom, are presented in a companion paper \citep{guha05b}.
For this work, we have assumed a distance
to M31 of 783\,kpc \citep[$(m - M)_0 = 24.47$; e.g.,][]{holl98,stan98}.

\section{Observations and Data Analysis}
\label{txt:obsdata}

SDSS \citep{york00} is an imaging and spectroscopic
survey that has mapped $\sim 1/4$
of the sky. Imaging data in the five SDSS bandpasses 
\citep[$u,g,r,i,z$;][]{fuku96,gunn98,hogg01,gunn05} are processed through pipelines
to measure photometric and astrometric properties
\citep{lupt99, stou02, smit02, pier03,ivez04,lupt05,tuck05} and
to select targets for spectroscopic followup. The SDSS scan used in this work
 was carried out on 5 October 2002 \citep[see][]{zuck04a}, processed with the same
 pipeline as Data Release 1 \citep{abaz03}, and spans $\sim 18^{\circ} \times 2.5^{\circ}$ along the major axis of M31.
For dereddening and conversion from SDSS magnitudes to
$V,I$ magnitudes, we use \citet*{schl98} and
\citet{smit02}, respectively.

\begin{figure}[th]
\plotone{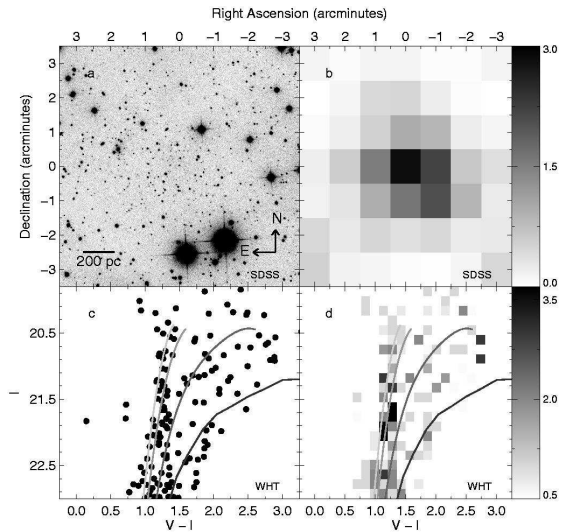}
\caption{\label{fig:sdss_wht}
Andromeda X as seen by SDSS and WHT: a) Sum of SDSS $g,r,i$ images centered on And X.
b) The spatial density of presumed SDSS ``blue'' red giant stars in the same area, binned $1\arcmin \times 1\arcmin$ and smoothed with a $1\farcm5$ FWHM Gaussian. The grayscale bar to the right indicates the significance of the spatial overdensity, in units of the background standard deviation.
c) Dereddened WHT CMD of stars within $2\arcmin$ of the center of And X, with fiducial sequences overplotted for Galactic globular clusters with metallicities of (left to right) [Fe/H] = $-2.2$ (M15), $-1.6$ (M2), $-0.7$ (47~Tuc), and $-0.3$ (NGC~6553) \citep{daco90,saga99}, shifted to a distance modulus of 24.47.
d) Dereddened WHT Hess diagram of stars within $2 \arcmin$ of the center of And X, minus  the area-scaled Hess diagram of a control field. The data are binned by $0\fm15$ in $(V-I)$ and $I$. The grayscale bar indicates the number of stars in each bin. Overplotted fiducials are as for c).
}
\vskip-0.5cm
\end{figure}

By selecting stars in the SDSS M31 scan with colors and magnitudes characteristic of
red giant branch (RGB) stars at the distance of M31, we discovered 
\aix, a new dwarf spheroidal satellite of M31 \citep{zuck04b}. Owing to their low metallicity \citep{harb05}, the RGB stars in \aix\ are relatively blue; limiting our photometric criteria to select {\em blue} RGB stars at the distance of M31, a large number of additional, less prominent overdensities were revealed. It is important to note that these stellar overdensities are not directly visible in SDSS images (see, e.g., Fig.\ref{fig:sdss_wht}a), and are only detected as enhancements of the 
spatial density of stars within a specific color-magnitude bin (Fig.\ref{fig:sdss_wht}b); in SDSS data, such enhancements may consist of 
a few stars spread across several square arcminutes. 
Deeper
 followup observations are therefore required in order to determine the nature of these overdensities, in particular to ascertain if they represent new 
M31 satellites.

In the autumn of 2004, we used the 4.2m William Herschel Telescope (WHT) and 2.5m Nordic Optical Telescope (NOT) on La Palma to image targets selected from these SDSS stellar overdensities. We observed at the WHT the nights of 14 and 15 September 2004, using the Prime Focus Imaging Camera (PFIP) with Harris $VI$ filters. Unfortunately, we were only able to obtain short integrations of one of the satellite candidates under suitable observing conditions (seeing $\lesssim 1.5\arcsec$). The WHT images were reduced with standard {\tt IRAF} tasks, and 
{\tt DAOPHOT II/ALLSTAR} \citep{stet94}
was used to obtain $V$ and $I$ photometry, which we bootstrap-calibrated with SDSS data.
Figure \ref{fig:sdss_wht}c shows the resulting dereddened $(I,V-I)$ color-magnitude diagram (CMD), for stars in a circular $2\arcmin$-radius region centered on the location of the satellite candidate. Despite significant field contamination, 
an RGB is visible. Subtracting the area-scaled CMD of an outer control region 
yields the Hess diagram in Figure \ref{fig:sdss_wht}d.
The narrowness, shape, and color of the RGB indicate the presence of a spatially distinct, metal-poor ([Fe/H] $\lesssim -1.6$) stellar population at roughly the distance of M31 (the tip of the RGB (TRGB) appears to be at $I \sim 20.2\pm0.2$), similar to that of \aix\, \citep{zuck04b,harb05}. The additional similarities between the observed candidate and \aix, with regard to low surface brightness (i.e., no obvious unresolved luminous component) and size ($\sim$ few arcminutes), are further indications that the object is
a new dwarf spheroidal (dSph) satellite of M31; we have therefore given it the name Andromeda X (\ax). 

\begin{figure}[th]
\plotone{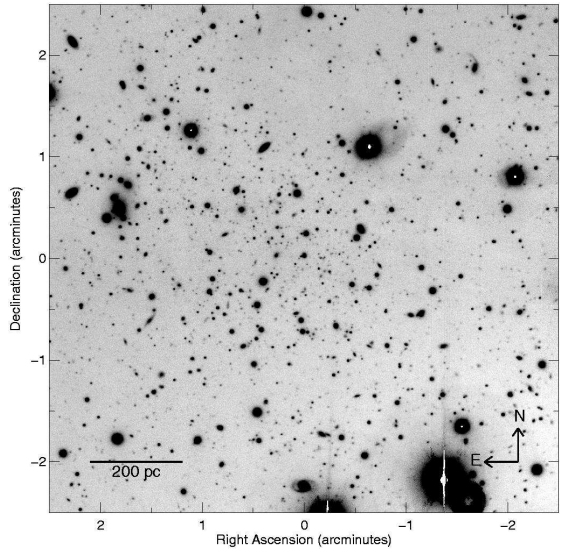}
\caption{\label{fig:notimg}
And X as seen by the NOT: summed $g,r,i$ images centered on And X.
}
\vskip -0.5cm
\end{figure}

\begin{figure}[th]
\epsscale{0.80}
\plotone{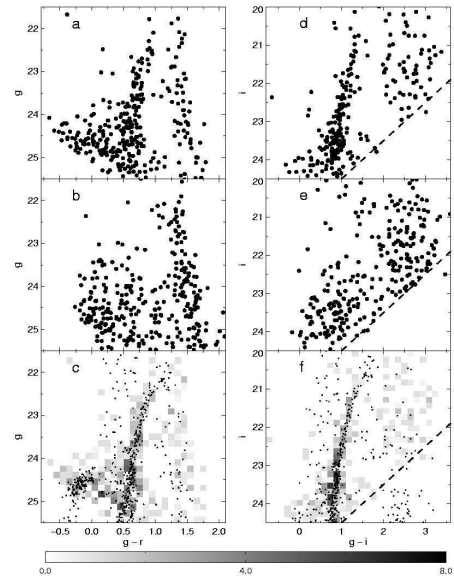}
\caption{\label{fig:notcmds} 
Dereddened CMDs of And X from NOT data: a) $(g,g-r)$ CMD of stars within $2\arcmin$ of the center of And X. b) $(g,g-r)$ CMD of stars from an outer control region.
c) $(g,g-r)$ Hess diagram of stars within $2\arcmin$ of the center of And X, minus the area-scaled Hess diagram of the control region. The data are binned by $0\fm15$ in $g$ and $(g-r)$. The SDSS CMD of an inner annulus of NGC 2419 (a Galactic globular cluster; [Fe/H]$ = -2.12$), dereddened and shifted to a distance modulus of 24.24 (see \S 3), is overplotted as dots.
d) $(i,g-i)$ CMD of stars within $2\arcmin$ of the center of And X. e) $(i,g-i)$ CMD of stars from the outer control region. f) $(i,g-i)$ Hess diagram of the central region, minus the scaled control region, binned by $0\fm15$ in $i$ and $0\fm10$ in $(g-i)$. The CMD of NGC 2419 is overplotted, as in c). The dashed line in d) - f) shows the approximate detection limit of our data; the grayscale bar indicates the number of stars per bin in c) and f).
}
\vskip -0.3cm
\end{figure}

During our subsequent run at the NOT (6 -- 11 October 2004), we used the Mosaic Camera (MOSCA) and SDSS $gri$ filters to obtain deeper imaging of this candidate with excellent seeing (typically $0\farcs6 - 0\farcs9$). 
We used the same reduction and analysis tasks 
to obtain SDSS-calibrated $gri$ photometry.
Figure \ref{fig:notimg} shows the coaddition of the NOT $g,r,i$ imaging data, in which \ax\, is revealed as an elliptical stellar overdensity (oriented roughly NE -- SW) at the center. The superior depth and resolution of the NOT data are evident in the resulting CMDs of the central $2\arcmin$-radius region (Fig. \ref{fig:notcmds}a and d), in which \ax's narrow, blue RGB clearly stands out. The background-subtracted dereddened Hess diagrams (Figs.\ref{fig:notcmds}c and f) are overplotted with the dereddened SDSS CMD of NGC~2419, a Galactic globular cluster ([Fe/H]$ = -2.12$), shifted to a distance modulus of 24.24 (see \S \ref{txt:props}); given the uncertainties in both \ax's distance and reddening, the agreement between \ax\, and NGC~2419 is remarkable, even hinting at the possible presence of blue horizontal branch stars in the $(g,g-r)$ CMD, and provides further evidence for \ax's low metallicity ([Fe/H] $\sim -2$).

\section{Properties of And X}

\label{txt:props}

Using background-subtracted photometric data from the NOT observations, we constructed a spatial density map of probable \ax\, members in radial bins of $15\arcsec$, to which we fit two-dimensional \citet{king66} and \citet{sers68} profiles.
We then applied the techniques of \citet{knia04} to the NOT $g,r,i$ images, masking foreground stars
and subtracting a fitted sky level, binning the resulting data in circular annuli to measure the central surface brightness and total magnitude in each filter. No unresolved luminous component was detected in \ax, so we used the method of \citet{zuck04a} to estimate the total surface brightness and luminosity by comparison with SDSS observations of the Pegasus dIrr; comparison to the luminosity function of the Draco dSph \citep{oden01} confirms that these estimates are reasonable for a range of stellar populations \citep{zuck04a}. The undereddened central surface brighnesses derived for \ax\, in this way are $27\fm6 \pm 0.4$, $26\fm7 \pm 0.3$, and $26\fm3 \pm 0.3$ in $g$, $r$, and $i$, respectively ($27\fm1 \pm 0.5$ in $V$); the measured total magnitudes in $g,r,i$ are $17\fm1 \pm 0.4$, $16\fm1 \pm 0.3$, and $15\fm8 \pm 0.3$ ($16\fm5 \pm 0.5$ in $V$). Table \ref{tbl:pars} lists the derived structural parameters and dereddened $V$-band magnitudes for 
\aix\, and \ax. 

Determining a more precise distance to \ax\, with the data in hand is problematic, due to the sparseness of \ax's CMD in the region of the TRGB. We closely examined the field-star-subtracted $I$-band stellar luminosity functions derived from both WHT and NOT data within the central $2\arcmin$ radius of \ax; however, even convolution with an edge detection kernel did not yield an unambiguous TRGB magnitude, with the measured 
$I_{TRGB}$ ranging from approximately 
$20.14$ to $20.36$, after correction for a foreground reddening of $A_I = 0.24$. Assuming a metallicity of [Fe/H]$ = -2.2$ and a TRGB color of $(V - I)_{TRGB} =1.2$ \citep[as for \aix;][]{zuck04b}, from the 
calibration of \citet{daco90} we obtain the relation $(m-M)_0 = I_{TRGB} + 3.98$, yielding a distance modulus range of $(m-M)_0 \sim 24.12$ to $24.34$, with uncertainties on the order of $0.1$ mag ($667$ kpc to $738$ kpc, $\pm 30$ to $35$ kpc).  At an M31 distance of 783 kpc, the angular separation of \ax\, from the center of M31, $\sim 5\fdg5$, gives a projected separation of $75$ kpc; considering the above range of \ax\, distance estimates puts \ax\, at $87$ kpc to $138$ kpc from 
the center of M31, well within M31's estimated virial radius \citep[e.g.,][]{bens02a} and the radius at which M31 halo stars have been found \citep{guha05a}. Thus \ax\, is in all likelihood a bound satellite of M31, with a dereddened integrated magnitude of $V_{\rm tot} \sim 16.1$, translating to an absolute magnitude of $M_V \sim -8.1 \pm 0\fm5$.

\begin{figure}[th]
\plotone{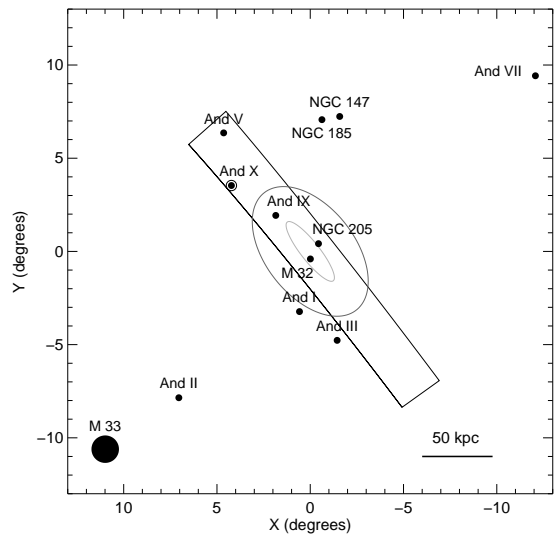}
\caption{\label{fig:m31sats} Projected distribution of M31's nearest satellites \citep[after][]{ferg02}, with the addition of And IX and And X. The inner ellipse indicates the rough optical size of M31's disk, and the outer ellipse the approximate extent of the INT survey \citep[see, e.g.,][]{ferg02}. The coverage of the SDSS M31 scan is shown by the long stripe along the major axis of M31.  
}
\vskip -0.5cm
\end{figure}

\begin{deluxetable}{lccc}
\tabletypesize{\small}
\tablecaption{Properties of And IX and And X  \label{tbl:pars}}
\tablewidth{0pt}
\tablehead{
\colhead{Parameter\tablenotemark{a}} &
\colhead{And IX} & \colhead{And X}
}
\startdata
R.A.(J2000.0)                    & ~~00 52 52.8     & 01 06 33.7  \\
DEC.(J2000.0)                    & $+$43 12 00      & $+$44 48 15.8  \\
King $r_c$				& $1\farcm33$\tablenotemark{b} & $1\farcm33$ \\
King $r_t$				& $5\farcm9$\tablenotemark{b} & $7\farcm2$ \\
S\'{e}rsic $n$				& $1.6$\tablenotemark{b} & $1.2$ \\
A$_{\rm V}$           & 0\fm26    & 0\fm41                       \\
$\mu_{\rm 0,V}$                  & $26\fm77 \pm 0\fm09$ & $26.7 \pm 0\fm5$\\ 
V$_{\rm tot}$                   & $16\fm17 \pm 0\fm06$ & $16.1 \pm 0\fm5$ \\ 
(m$-$M)$_0$                      & $24\fm48 \pm 0\fm20$ & $24\fm12 -  24\fm34 \pm 0\fm10$\\
M$_{\rm tot,V}$                & $-8\fm3$ & $-8\fm1$ 
\enddata
\tablenotetext{a}{Surface brightnesses and integrated magnitudes are corrected for the mean
Galactic foreground reddenings, A$_{\rm V}$, shown. Unless otherwise noted, values for And~IX are from \citet{zuck04b}.}
\tablenotetext{b}{From \citet{harb05}.}
\end{deluxetable}

\section{Discussion}

As shown in Table \ref{tbl:pars}, \ax\, is comparable in size, surface brightness, and apparent magnitude to \aix, and, despite the uncertainty in its distance, \ax\, appears to be 
somewhat lower in luminosity ($M_V \sim -8.1$). We conclude that \ax\, is a new extremely low luminosity dSph companion of M31, a result confirmed by kinematic data \citep{guha05b}. The earlier discovery of \aix\, raised the question of whether such low luminosity galaxies were a rarity in the Local Group, or if \aix\, represented the tip of an iceberg, one of a large population of faint  dwarf satellites that have 
remained undetected owing to their extremely low surface brightnesses and low luminosities. \ax's discovery,
in conjunction with the recent discovery of the Ursa Major dwarf \citep{will05}, strongly suggests that the latter scenario is closer to the truth -- that M31 and the Milky Way may in fact have a large number of low luminosity satellites.

Further analysis of SDSS data
is therefore likely to yield more such dwarfs. To test this scenario, we are following up other stellar overdensities in the SDSS M31 scan with deeper observations on other telescopes.  Intriguingly,  in projection \aix\, and \ax\, lie relatively close to each other and to the major axis of M31, northeast of M31's center (Fig. \ref{fig:m31sats}); as more new M31 satellites are found (as appears quite likely), we will be able to place constraints on the spatial distribution of M31 satellites and their (an)isotropy \citep[e.g.,][]{koch05}, as well as on the satellite luminosity function. Thus the discovery and characterization of extremely low luminosity galaxies like \aix\, and \ax\, play a critical role in reconciling the predictions of current CDM galaxy formation models with the observable local Universe.

\acknowledgements

DBZ acknowledges support from a National Science Foundation International Postdoctoral Fellowship.
EFB acknowledges the financial support provided through the European
Community's Human Potential Program under contract
HPRN-CT-2002-00316, SISCO. We thank M. Schirmer and T. Pursimo for assistance in reducing WHT and NOT data, respectively. This work utilized observations made with the Nordic Optical Telescope, operated on the island of La Palma jointly by Denmark, Finland, Iceland, Norway, and Sweden, in the Spanish Observatorio del Roque de los
Muchachos of the Instituto de Astrof\'{i}sica de Canarias;  
these observations were funded by the
Optical Infrared Coordination Network (OPTICON),
a major international collaboration supported by the
Research Infrastructures Programme of the
European Commission's Sixth Framework Programme.

Funding for the creation and distribution of the SDSS Archive has been
provided by the Alfred P. Sloan Foundation, the Participating
Institutions, the National Aeronautics and Space Administration, the
National Science Foundation, the U.S. Department of Energy, the
Japanese Monbukagakusho, and the Max Planck Society.  The SDSS Web
site is \texttt{http://www.sdss.org/}.

The SDSS is managed by the Astrophysical Research Consortium (ARC) for the Participating Institutions. The Participating Institutions are The University of Chicago, Fermilab, the Institute for Advanced Study, the Japan Participation Group, The Johns Hopkins University, the Korean Scientist Group, Los Alamos National Laboratory, the Max-Planck-Institute for Astronomy (MPIA), the Max-Planck-Institute for Astrophysics (MPA), New Mexico State University, University of Pittsburgh, University of Portsmouth, Princeton University, the United States Naval Observatory, and the University of Washington.


\begin{thebibliography}{}

\bibitem[Abazajian et al.(2003)]{abaz03}
        Abazajian, K., et al. 2003, \aj, 126, 2081



\bibitem[Benson et al.(2002a)]{bens02a} Benson, A.~J., Lacey, 
C.~G., Baugh, C.~M., Cole, S., \& Frenk, C.~S.\ 2002a, \mnras, 333, 156 

\bibitem[Benson et al.(2002b)]{bens02b} Benson, A.~J., Frenk, 
C.~S., Lacey, C.~G., Baugh, C.~M., \& Cole, S.\ 2002b, \mnras, 333, 177 



\bibitem[Da Costa \& Armandroff(1990)]{daco90}
        Da Costa, G.\ S., \& Armandroff, T.\ E. 1990, \aj, 100, 162

\bibitem[Ferguson et al.(2002)]{ferg02}
        Ferguson, A.\ M.\ N., Irwin, M.\ J., Ibata, R.\ A., Lewis, G.\ F., \&
        Tanvir, N.\ R. 2002, \aj, 124, 1452

\bibitem[Fukugita et al.(1996)]{fuku96}
        Fukugita, M., Ichikawa, T., Gunn, J.\ E., Doi, M., Shimasaku, K., \&
        Schneider, D. P. 1996, \aj, 111, 1748


\bibitem[Guhathakurta et al.(2005a)]{guha05a}

	Guhathakurta, P., et al. 2005a, \nat, {\em submitted} 

\bibitem[Guhathakurta et al.(2005b)]{guha05b}
	Guhathakurta, P. et al. 2005b, {\em in preparation}


\bibitem[Gunn et al.\ (1998)]{gunn98} Gunn, J.E. et al. 1998, \aj, 116, 3040

\bibitem[Gunn et al.(2005)]{gunn05} Gunn, J.E. et al. 2005, \aj, {\em submitted}

\bibitem[Harbeck et al.(2005)]{harb05} Harbeck, D., Gallagher, J., Grebel, E., Koch, A., \& Zucker, D. 2005, \apj, 623, 159


\bibitem[Hogg et al.(2001)]{hogg01}
        Hogg, D.W., Finkbeiner, D.P., Schlegel, D.J., and Gunn, J.E. 2001,
        \aj, 122, 2129

\bibitem[Holland(1998)]{holl98} Holland, S.\ 1998, \aj, 115, 1916


\bibitem[Ivezi\'{c} et al.(2004)]{ivez04} Ivezi\'{c}, \v{Z}. et al., AN, 325, 583

\bibitem[King(1966)]{king66} King, I.~R.\ 1966, \aj, 71, 64 


\bibitem[Kniazev et al.(2004)]{knia04}
        Kniazev, A.\ Y., Grebel, E.\ K., Pustilnik, S.\ A., Pramskij, A.\ G.,
        Kniazeva, T.\ F., Prada, F., \& Harbeck, D. 2004, \aj, 127, 704


\bibitem[Klypin et al.(1999)]{klyp99}
        Klypin, A., Kravtsov, A.\ V., Valenzuela, O., \& Prada, F. 1999, \apj, 522, 82

\bibitem[Koch \& Grebel(2005)]{koch05}
	Koch, A., \& Grebel, E. 2005, \aj, {\em in press} ({\texttt astro-ph/0509258})

\bibitem[Kravtsov, Gnedin, \& Klypin(2004)]{krav04}
        Kravtsov, A., Gnedin, O., \& Klypin, A. 2004, \apj, 609, 482



\bibitem[Lupton(2005)]{lupt05} Lupton, R.\ H. 2005, \aj, {\em submitted}

\bibitem[Lupton, Gunn, \& Szalay(1999)]{lupt99} Lupton, R., Gunn, J., 
	\& Szalay, A. 1999, \aj, 118, 1406




\bibitem[Moore et al.(1999)]{moor99}
        Moore, B., Ghigna, S., Governato, F., Lake, G., Quinn, T., Stadel, J., \& Tozzi, P. 1999, \apjl, 524, 19


\bibitem[Odenkirchen et al.(2001)]{oden01}
	Odenkirchen, M. et al. 2001, \aj, 122, 2538

\bibitem[Pier et al.(2003)]{pier03}
        Pier, J.R., Munn, J.A., Hindsley, R.B, Hennessy, G.S., Kent, S.M.,
        Lupton, R.H., \& Ivezic, Z. 2003, \aj, 125, 1559


\bibitem[Sagar et al.(1999)]{saga99}
        Sagar, R., Subramaniam, A., Richtler, T., \& 
        Grebel, E.\ K. 1999, \aaps, 135, 391


\bibitem[Schlegel et al.(1998)Schlegel, Finkbeiner, \& Davis]{schl98} Schlegel, D.\ J.,
        Finkbeiner, D.\ P., \& Davis, M.\ 1998, \apj, 500, 525

\bibitem[S\'{e}rsic(1968)]{sers68}
	S\'{e}rsic, J. 1968, Atlas de galaxias australes, C\'{o}rdova: Observatorio Astron\'{o}mico

\bibitem[Smith et al.(2002)]{smit02}
        Smith, J.A. et al. 2002, \aj, 123, 2121

\bibitem[Somerville(2002)]{some02} Somerville, R.~S.\ 2002, 
\apjl, 572, L23 

\bibitem[Stanek \& Garnavich(1998)]{stan98} Stanek, K., \& Garnavich, P.\ 1998, \apj, 503, L131

\bibitem[Stetson(1994)]{stet94} Stetson, P.~B.\ 1994, \pasp, 
106, 250 


\bibitem[Stoehr et al.(2002)]{stoe02} 
Stoehr, F., White, S.~D.~M., Tormen, G., \& Springel, V.\ 2002, \mnras, 
335, L84 



\bibitem[Stoughton et al.(2002)]{stou02}
        Stoughton, C. et al. 2002, \aj, 123, 485

\bibitem[Tucker et al.(2005)]{tuck05} Tucker, D., et al. 2005, \aj, {\em submitted}

\bibitem[van den Bergh(1999)]{vand99}
        van den Bergh, S. 1999, \aapr, 9, 273

\bibitem[Willman et al.(2002)]{will02} Willman, B.,
Dalcanton, J., Ivezi{\' c}, {\v Z}., Jackson, T., Lupton, R.,
Brinkmann, J., Hennessy, G., \& Hindsley, R.\ 
2002, \aj, 123, 848 

\bibitem[Willman et al.(2005)]{will05} Willman, B., et al.\, 2005, \apjl, 626, L85


\bibitem[York et al.(2000)]{york00} York, D.G.
        et al.\, 2000, \aj, 120, 1579

\bibitem[Zucker et al.(2004a)]{zuck04a}
        Zucker et al.\, 2004a, \apjl, 612, L117

\bibitem[Zucker et al.(2004b)]{zuck04b}
        Zucker et al.\, 2004b, \apjl, 612, L121

\end{thebibliography}
\end{document}